\documentclass[%
 reprint,
superscriptaddress,
 amsmath,amssymb,
 aps,
 prl,
]{revtex4-2}

\usepackage{textcomp}
\usepackage{graphicx}
\usepackage{dcolumn}
\usepackage{bm}

\graphicspath{{images/}}

\begin{document}

\preprint{APS/123-QED}

\title{Nonreciprocal Dzyaloshinskii-Moriya magnetoacoustic waves}

\author{M. K\"u\ss{}}
 \email{matthias.kuess@physik.uni-augsburg.de.}
 \affiliation{Experimental Physics I, Institute of Physics, University of Augsburg, 86135 Augsburg, Germany\looseness=-1}
 
\author{M. Heigl}
 \affiliation{Experimental Physics IV, Institute of Physics, University of Augsburg, 86135 Augsburg, Germany\looseness=-1}

\author{L. Flacke}
 \affiliation{Walther-Meißner-Institut, Bayerische Akademie der Wissenschaften, 85748 Garching, Germany}
 \affiliation{Physics-Department, Technical University Munich, 85748 Garching, Germany}

\author{A. H\"orner}
 \affiliation{Experimental Physics I, Institute of Physics, University of Augsburg, 86135 Augsburg, Germany\looseness=-1}

\author{M. Weiler}
 \altaffiliation{Present address: Fachbereich Physik, Technische Universität Kaiserslautern, 67663 Kaiserslautern, Germany}
 \affiliation{Walther-Meißner-Institut, Bayerische Akademie der Wissenschaften, 85748 Garching, Germany}
 \affiliation{Physics-Department, Technical University Munich, 85748 Garching, Germany}

\author{M. Albrecht}
 \affiliation{Experimental Physics IV, Institute of Physics, University of Augsburg, 86135 Augsburg, Germany\looseness=-1}
 
\author{A. Wixforth}
 \affiliation{Experimental Physics I, Institute of Physics, University of Augsburg, 86135 Augsburg, Germany\looseness=-1}


\date{\today}

\begin{abstract}
We study the interaction of surface acoustic waves with spin waves in ultra-thin CoFeB/Pt bilayers. Due to the interfacial Dzyaloshinskii-Moriya interaction (DMI), the spin wave dispersion is non-degenerate for oppositely propagating spin waves in CoFeB/Pt. In combination with the additional nonreciprocity of the magnetoacoustic coupling itself, which is independent of the DMI, highly nonreciprocal acoustic wave transmission through the magnetic film is observed. We systematically characterize the magnetoacoustic wave propagation in a thickness series of CoFeB($d$)/Pt samples as a function of magnetic field magnitude and direction, and at frequencies up to 7~GHz. We quantitatively model our results to extract the strength of the DMI and magnetoacoustic driving fields.
\end{abstract}

\maketitle

Surface acoustic waves (SAW) have made their way into both technology and research over the last few decades. They are easily excited and detected on piezoelectric crystals and have many different applications, most notably for rf signal processing in telecommunications~\cite{Campbell.1998} but also as sensors~\cite{Paschke.2017} and even microfluidic lab-on-a-chip devices~\cite{ZenoGuttenberg.2005}. 
Here, we would like to report on a striking observation if SAW and spin waves (SW) in a magnetic thin film are coupled: The Dzyaloshinskii-Moriya interaction (DMI)~\cite{Moriya.1960, Dzyaloshinsky.1958} leads to a pronounced nonreciprocal behavior for the SAW/magnetic thin film hybrid. As SAWs on piezoelectrics are usually propagating reciprocally, i.e., their properties do not depend on the direction of propagation along a specific crystal axis, the possibility of nonreciprocity would open completely new fields of applications, ranging from acoustic diodes~\cite{Liang.2009} to chiral phononics~\cite{EvenThingstad.2019}. Over the years, various mechanisms breaking the reciprocity of SAW propagation have so far been investigated both experimentally as well as in theory~\cite{Emtage.1976, Heil.1982, Rotter.1999, Rotter.1999b}. For instance, the acoustoelectric amplification~\cite{Rotter.1999b} has been addressed before. But also the nonreciprocal interaction of SAWs with a magnetic medium has been investigated theoretically a long time ago~\cite{Emtage.1976}.

In this letter, we experimentally demonstrate the strong coupling between SAWs and SWs in thin magnetic films with DMI to generate highly nonreciprocal magnetoacoustic surface waves (MASW). We show that this interaction leads to an additional nonreciprocity effect in frequency space, extending well beyond the know amplitude nonreciprocity~\cite{Dreher.2012, Sasaki.2017, Tateno.2020} being always present in magnetic media due to the elliptical polarization of SAWs and SWs. In total, the coupling of SAWs with SWs results in double nonreciprocal MASWs with a nonreciprocal contrast of up to 27.9~dB/mm at 6.77~GHz.
We fit our experimental results to a modified Landau-Lifshitz-Gilbert model to extract the strength of the DMI and magnetoacoustic driving fields. The DMI is in full agreement with previous reports and the magnetoacoustic driving fields are quantitatively accounted for by combined magnetoelastic and magnetorotation~\cite{Maekawa.1976} coupling.

The DMI at a ferromagnetic – heavy metal interface energetically favors spin structures with a fixed chirality, which is fundamentally responsible for the formation of magnetic skyrmions~\cite{Muhlbauer.2009}. Counter-propagating SWs have opposite spatial chirality and are thus non-degenerate in the presence of DMI~\cite{Nembach.2015, Lee.2016}. The potential of nonreciprocal DMI MASWs has been theoretically discussed by Verba et al.~\cite{Verba.2018}.  While in Ref.~\cite{Verba.2018}, Ni/Pt thin film bilayers have been suggested, here we employ CoFeB/Pt bilayers, which have lower magnetic damping~\cite{JWalowski.2008, A.Conca.2013}.
\begin{figure}[b]
\includegraphics[width = 0.48\textwidth]{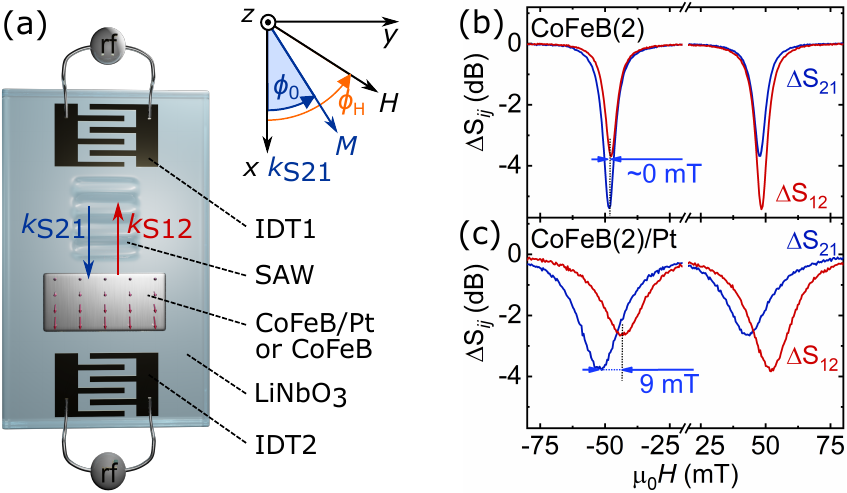}
\caption{
\label{fig:1}
(a) Schematic illustration of the experimental setup and the coordinate system. A scaled drawing is shown in the Supplemental Material~\cite{Suppl:Ref}. Nonreciprocal MASWs are characterized by different transmission amplitudes $\Delta S_{21}$ and $\Delta S_{12}$ for oppositely propagating SAW $k_{S21}$ and $k_{S12}$. The SAW transmission curves of (b) CoFeB(2) and (c) CoFeB(2)/Pt are obtained at 6.88~GHz, 6.90~GHz and $\phi_\text{H} = 64.8^\circ$. Only CoFeB(2)/Pt reveals a nonreciprocity in both transmission amplitude at resonance and resonant magnetic field.
}
\end{figure}

We study acousto-magnetic devices as schematically shown in FIG.~\ref{fig:1}(a). Details about the sample preparation are presented in the Supplemental Material~\cite{Suppl:Ref}. For all samples, we measured the saturation magnetization $M_\text{s}$ of the Co$_{40}$Fe$_{40}$B$_{20}$ layers by superconducting quantum interference device-vibrating sample magnetometry, see TABLE SIII~\cite{Suppl:Ref}. The determined $M_\text{s}$ values for LiNbO$_3$/CoFeB(2) and LiNbO$_3$/CoFeB(1.4 - 2.0)/Pt(3) (numbers are the nominal thicknesses in nm) are in good agreement with literature~\cite{Cho.2013b}. 
The thicker samples were fabricated in a second sputter run and show higher values for $M_\text{s}$. The effective magnetization $M_\text{eff}$, Gilbert damping and inhomogeneous line broadening were determined by broadband ferromagnetic resonance measurements (FMR)~\cite{Suppl:Ref}.
A vector network analyzer was used to measure the SAW transmission of our delay lines utilizing a time-domain gating technique~\cite{Hiebel.2011, Dreher.2012}. We study MASWs up to $f=7.0$~GHz, by exploiting the 7th harmonic resonance frequency of the interdigital transducers (IDT). If the excitation and detection IDTs are interchanged, the propagation direction of the acoustic wave and the excited MASW with the wavevectors $k_{S21}>0$ or $k_{S12}<0$ is reversed, probing nonreciprocal effects.

In the following, we use the coordinate system shown in FIG.~\ref{fig:1}(a). The angle $\phi_\text{H}$ ($\phi_0$) defines the direction of the external magnetic field $\mathbf{H}$ (static magnetization $\mathbf{M}$).  FIG.~\ref{fig:1}(b, c) depict the SAW transmission $\Delta S_{21}$, $\Delta S_{12}$ of the CoFeB(2) and CoFeB(2)/Pt samples at $6.9$~GHz and $\phi_\text{H} = 64.8^\circ$.
We define the relative change of the background corrected SAW transmission signal as $\Delta S_{ij} (\mu_0 H) = S_{ij} (\mu_0 H) - S_{ij} (-200~\text{mT})$, where $\Delta S_{ij}$ is the magnitude of the complex transmission signal with $ij=21,12$.
Clearly, a large difference $\Delta S_{21} \neq \Delta S_{12}$ is observed in FIG.~\ref{fig:1}(b,c), corresponding to nonreciprocal MASW propagation.
Two independent mechanisms lead to this nonreciprocity. First, the helicity mismatch between the magnetoacoustic driving field and magnetization precession~\cite{Dreher.2012, Sasaki.2017, Tateno.2020}, that induces different transmission magnitude at resonance in both samples. Second, the DMI, which causes resonance field shifts of 9~mT only in CoFeB(2)/Pt. Both effects will be discussed in more detail.

The Rayleigh type SAW generates strain $\epsilon_{xx} (x,t), \epsilon_{zz} (x,t)$, and $\epsilon_{xz} (x,t)$ in the magnetic film, with the frequency $f$, the periodicity $\lambda=c_\text{SAW}/f$ and the wavevector $|k|=2\pi/\lambda$, that are given by the phase velocity of a metalized LiNbO$_3$ surface $c_\text{SAW}$~\cite{Morgan.2007}. The shear strain $\epsilon_{xz} (x,t)$ is phase shifted by $90^\circ$ with respect to $\epsilon_{xx} (x,t)$. For a polycrystalline magnetic film, the complex amplitudes of the SAW induced in-plane $h_\text{ip}$ and out-of-plane $h_\text{oop}$ field components that potentially excite SWs are
\begin{align}
 &\mu_0 h_\text{ip\hphantom{p}} = 2 b_{xx} |k| |u_{z,0}| \sin{\phi_0} \cos{\phi_0} \nonumber\\
 &\mu_0 h_\text{oop} = 2 b_{xz} |k| |u_{z,0}| \cos{\phi_0}.
 \label{eq:1}
\end{align}
For magnetoelastic coupling, $b_{ij} = b_{1,2} \tilde{a}_{ij}$ with $ij=xx,xz$, where $b_1=b_2$ are the magnetoelastic coupling constants for polycrystalline CoFeB. We determine \mbox{$\tilde{a}_{ij} = \frac{\epsilon_{ij,0}}{|k| |u_{z,0}|}$} with a finite element method simulation, as shown in the Supplemental Material~\cite{Suppl:Ref}. Furthermore, $u_{z,0}$ is the amplitude of the lattice displacement in the z-direction.

A recent, related study~\cite{Xu.2020} reported that magnetorotational coupling~\cite{Maekawa.1976} induces additional driving fields with the same symmetry as the shear strain magnetoelastic driving field $h_\text{oop}$. The additive contribution due to magnetorotational coupling is expressed in Eq.~(\ref{eq:1}) by $b_{xz} = -B_\text{u} \frac{\omega_{xz,0}}{|k| |u_{z,0}|}$ with the uniaxial effective out-of-plane anisotropy $B_\text{u}=-\frac{1}{2} \mu_0 M_\text{eff}$. Here $\omega_{xz,0}$ is the amplitude of the rotation tensor element $\omega_{xz}= \frac{1}{2} \left( \frac{\partial u_\text{x}}{\partial z} - \frac{\partial u_\text{z}}{\partial x} \right) $.

Following the approach by Dreher et al.~\cite{Dreher.2012}, and additionally considering the effects of dipolar interactions and DMI~\cite{Moon.2013}, we obtain the anisotropic SW dispersion
\begin{equation}
    f = \frac{ \mu_0 \gamma}{2 \pi} \sqrt{H_{11} H_{22}}
    -\frac{\gamma}{\pi M_\text{s}} D_\text{eff} \, k \sin(\phi_0),
    \label{eq:2}
\end{equation}

where we assume that $\bf{M} \parallel \bf{H}$ ($\phi_0 = \phi_\text{H}$). In Eq.~(\ref{eq:2}), $\gamma$ is the gyromagnetic ratio and $H_{11}$, $H_{22}$ are terms, depending on the external magnetic field, the exchange field, the in-plane and out-of-plane magnetic anisotropy fields and dipolar fields~\cite{Suppl:Ref}.
\begin{figure}[b]
\includegraphics{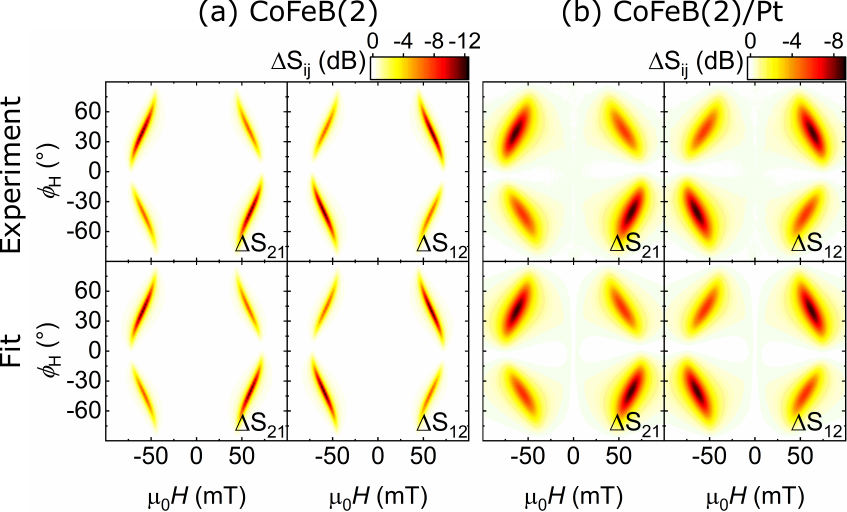}
\caption{
\label{fig:2}
Change of the SAW transmission  $\Delta S_{ij}$  as a function of the orientation and magnitude of the external magnetic field. The experimental $\Delta S_{21}$ and $\Delta S_{12}$ of the (a) CoFeB(2) single layer and (b) CoFeB(2)/Pt bilayer (upper row) are measured at the resonance frequencies of the SAW delay lines at 6.90~GHz and 6.88~GHz, respectively. Both samples reveal nonreciprocal behavior with respect to the transmission magnitude. An additional nonreciprocal shift of the resonance fields is induced by DMI in the CoFeB(2)/Pt sample. Numerical fits (lower row).
}
\end{figure}
The final term in Eq.~(\ref{eq:2}) results from the DMI, here being parametrized by the thickness-averaged effective DMI constant $D_\text{eff}$. The DMI causes the resonant-field nonreciprocity of the MASW, as observed in FIG.~\ref{fig:1}(c). The second mechanism which induces nonreciprocal MASWs is caused by a SAW-SW helicity mismatch.
If the propagation direction of the SAW is inverted, the helicity of the SAW and thus of the driving fields are reversed.  Due to the fixed helicity of the magnetization precession, this results in nonreciprocal coupling efficiency~\cite{Dreher.2012, Sasaki.2017, Tateno.2020}.

To investigate the two nonreciprocal mechanisms in more detail, we perform measurements such as being shown in FIG.~\ref{fig:1}(b,c) as a function of $\phi_\text{H}$. The upper row of FIG.~\ref{fig:2} depicts the experimentally determined $\Delta S_{ij}$, obtained for delay lines loaded with CoFeB(2) and CoFeB(2)/Pt films and operated at 6.9~GHz. FIG.~\ref{fig:2} reveals the expected fourfold symmetry for magnetoacoustically driven resonance~\cite{Weiler.2011}. Furthermore, we find that the linewidth of the SW resonance of the CoFeB(2)/Pt sample is larger than that of the CoFeB(2) device. This can be attributed to spin pumping~\cite{Tserkovnyak.2002}. The fourfold symmetry is obviously broken for both samples, as the $\Delta S_{21}$ transmission curves show a more intense SW resonance at $H<0$, $+45^\circ$ than at $H>0$, $+45^\circ$. For the opposite propagation direction $\Delta S_{12}$, this asymmetry is reversed. This nonreciprocity is a consequence of the SAW-SW helicity mismatch. Moreover, in contrast to the CoFeB(2) sample, the resonance fields of the CoFeB(2)/Pt sample are nonreciprocally shifted. According to Eq.~(\ref{eq:2}), the additional DMI contribution lifts the degeneracy of the resonant magnetic fields for counter-propagating SAWs proportional to $\sin \phi_0$.

In the lower row of FIG.~\ref{fig:2}, we show results of a fit to our $\Delta S_{ij}$ data.
The fit function is derived from the magnetic susceptibility taking dipolar, exchange, and DMI into account. Additionally, the symmetry of the driving fields and the field drag effect ($\phi_0 \neq \phi_\text{H}$) are considered. We included exponentially decaying driving fields $h_\text{ip} (x)$ and $h_\text{oop} (x)$ along the SAW propagation direction. More details about the fitting procedure and derivation of the equations are found in the Supplemental Material~\cite{Suppl:Ref}. 
The fit parameters are the uniaxial in-plane and surface out-of-plane magnetic anisotropy fields, $b_{xx}$ and $b_{xz}$, the SW damping constant and $D_\text{eff}$.
We find overall excellent agreement of experiment and fit with reasonable fit parameters summarized in the supplemental TABLE~SIII~\cite{Suppl:Ref}.

Furthermore, the extracted damping and effective magnetization are in agreement with FMR measurements performed on reference samples~\cite{Suppl:Ref}. 
The good agreement of the effective damping constants is in contrast to previous findings~\cite{Dreher.2012, Weiler.2011, Gowtham.2015} and further demonstrates that our phenomenological model is adequate.

\begin{figure}[b]
\includegraphics{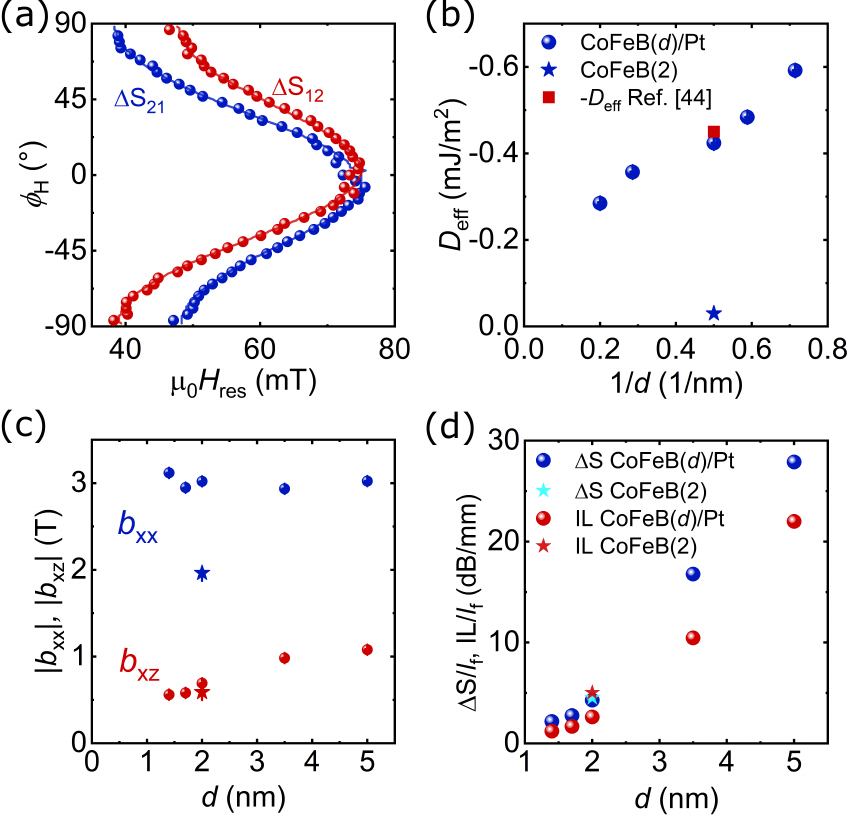}
\caption{
\label{fig:3}
(a) Angular dependence of the resonance fields of the CoFeB(2)/Pt sample in FIG.~\ref{fig:2}. Dots denote the experimental data, solid curves the fit. (b)~The magnitude of the effective DMI constant as a function of the inverse of the film thickness agrees with the expected linear behavior, revealing the interface character of the underlying interaction. (c)~The driving field fit parameters $b_{xx}$ and $b_{xz}$ for the CoFeB($d$)/Pt thickness series (dots) and the CoFeB(2) sample (stars). The error bars in (b) and (c) are smaller than the symbol size. (d)~Summary of the highest attained transmission nonreciprocity $\Delta S$ and the corresponding attenuation in the acoustic diode forward direction $IL$.
}
\end{figure}

We now focus on the nonreciprocity of the MASW propagation. A hypothetical, perfectly nonreciprocal SAW device would show 100\% transmission (zero loss) in the forward direction and no transmission ($\infty$ loss) in the reverse direction. Although the DMI induced splitting of the resonance fields $\Delta(\mu_0 H_\text{res})$ is up to 10~mT in the CoFeB(2)/Pt film, the attenuation dips in $\Delta S_{21}$ and $\Delta S_{12}$ overlap, as already shown in FIG.~\ref{fig:1}(c).
To study $\Delta(\mu_0 H_\text{res})$ in more detail, we plot the resonance fields $\mu_0 H_\text{res}$ as a function of $\phi_\text{H}$ in FIG.~\ref{fig:3}(a) for CoFeB(2)/Pt.
As expected from Eq.~(\ref{eq:2}), $\Delta(\mu_0 H_\text{res})$ is proportional to $\sin \phi_0$ and vanishes for the CoFeB sample without Pt, as shown in the Supplemental Material~\cite{Suppl:Ref}. The lines in Fig.~\ref{fig:3}(a) are the transmission minima obtained from the global fit results.

The DMI induced nonreciprocity of the MASW is given by the magnitude of the effective DMI constant $D_\text{eff}$, which is shown in FIG.~\ref{fig:3}(b) as a function of CoFeB layer thickness $d$. Brillouin light scattering (BLS) measurements of a similar CoFeB(2)/Pt film result in \mbox{$D_\text{eff}=0.45$~mJ/m$^2$}~\cite{Tacchi.2017}. This magnitude is in good accordance with our result of \mbox{$D_\text{eff}=-(0.424 \pm 0.001)$~mJ/m$^2$}~\cite{Suppl:Ref}. Due to the interfacial origin of the DMI, $D_\text{eff}$ in FIG.~\ref{fig:3}(b) is expected to be linearly proportional to the inverse of the film thickness and to vanish in the limit of infinitely thick films. We attribute the slight deviation from this linear proportionality, as being observed in FIG.~\ref{fig:3}(b), dominantly to the different sputter runs in which samples for $1/d > 0.4$~nm$^{-1}$ and $1/d < 0.4$~nm$^{-1}$ were fabricated. 
The effect of asymmetric dipolar fields which can also serve as a source of nonreciprocal SWs increases with film thickness~\cite{Gladii.2016}. Since this effect is caused by surface anisotropy fields, being affected by the different sputter runs, asymmetric dipolar fields may slightly contribute to the observed nonlinearity, especially for the thicker films ($d \geqq 3.5$~nm).

The nonreciprocity of the MASW, caused by the SAW-SW helicity mismatch depends on the ratio of $h_\text{oop}$ to $h_\text{ip}$ and thus $b_{xz}$ and $b_{xx}$, which are shown in FIG.~\ref{fig:3}(c).
With $\tilde{a}_{xx} = 0.49 \pm 0.1$ from the finite element study~\cite{Suppl:Ref}, we obtain for the magnetoelastic coupling constant of the CoFeB(2) film $b_1 = -(4.0\pm0.8)$~T. This is in good agreement with the literature value $b_1 = -3.8$~T~\cite{b1value_ref}\nocite{P.G.Gowtham.2016}. 
Interestingly, for the CoFeB(d)/Pt samples $b_1$ is increased by the Pt layer and we obtain $b_1 = -(6.5 \pm 1.7)$~T.
In contrast to $b_{xx}$, which is approximately constant for the CoFeB($d$)/Pt series, $b_{xz}$ increases with $d$. 
In the Supplemental Material~\cite{Suppl:Ref} we calculate the expected $b_{xz}$ due to magnetorotation and magnetoelastic contributions. Both mechanisms qualitatively reproduce the increase of $b_{xz}$ with $d$.
The contribution caused by magnetorotational coupling is a factor 1 to 2 higher than the magnetoelastic counterpart, but both effects individually underestimate $b_{xz}$ by a factor 2 to 3. We conclude that both mechanisms are present, add up constructively and thus quantitatively reproduce the observed $b_{xz}$. Consequently, the strong SAW-SW helicity mismatch effect is induced by both mechanisms.

In FIG.~\ref{fig:3}(d), we show for all samples the largest transmission nonreciprocity \mbox{$\Delta S = \text{max}(\Delta S_{21}(H,\phi_\text{H}) - \Delta S_{12}(H,\phi_\text{H}))$}, observed at $H = H^{\Delta S} ,\phi_\text{H} = \phi_\text{H}^{\Delta S}$, $f \approx 6.9$~GHz and normalized to the magnetic film length $l_\text{f}$. Maximum nonreciprocity is found for all samples in a range of $\phi_\text{H}^{\Delta S} = (33,...,45)^\circ$ and at $\mu_0 H^{\Delta S} = (21,...,84)$~mT, close to the resonant fields. 
The increase of $\Delta S$ with $d$ is caused by a combination of increasing SAW-SW helicity mismatch nonreciprocity (FIG.~\ref{fig:3}(c)), decreasing linewidth (FIG.~S6(a)~\cite{Suppl:Ref}), which partly compensates decreasing nonreciprocity induced by decreasing $D_\text{eff}$ (FIG.~\ref{fig:3}(b)), and increased magnetic film volume, which increases the magnitude of both nonreciprocal effects~\cite{Suppl:Ref}.
The CoFeB(5)/Pt sample shows the highest transmission nonreciprocity of 27.9~dB/mm, which is considerably larger than reported values for LiNbO$_3$/Ni of $\Delta S < $ 0.1~dB/mm~\cite{Ni_nonrecip}\nocite{Tateno.2020} and for GaAs/Fe$_3$Si of $\Delta S < $ 0.9~dB/mm~\cite{FeSi_nonrecip}\nocite{A.HernandezMinguez.2020}.
In contrast to a perfect acoustic diode with 100\% transmission in the forward direction, the transmission nonreciprocity $\Delta S$ in FIG.~\ref{fig:3}(d) comes along with an insertion loss $IL = -\Delta S_{21}(H^{\Delta S}, \phi_\text{H}^{\Delta S})$, which is caused by the SAW-SW interaction. Although the nonreciprocity $\Delta S$ of the CoFeB(2) and the CoFeB(2)/Pt sample is similar, the DMI permits a reduction of the insertion loss due to shifted resonance fields of counter-propagating SWs. In general, we observe $\Delta S > IL$ for all samples with Pt while $\Delta S < IL$ for CoFeB without Pt, demonstrating that the DMI plays an important role in optimizing the diode like behavior. The insertion loss could be further lowered by increasing $D_\text{eff}$ or by decreasing the SW damping.

\begin{figure}[b]
\includegraphics[width = 0.48\textwidth]{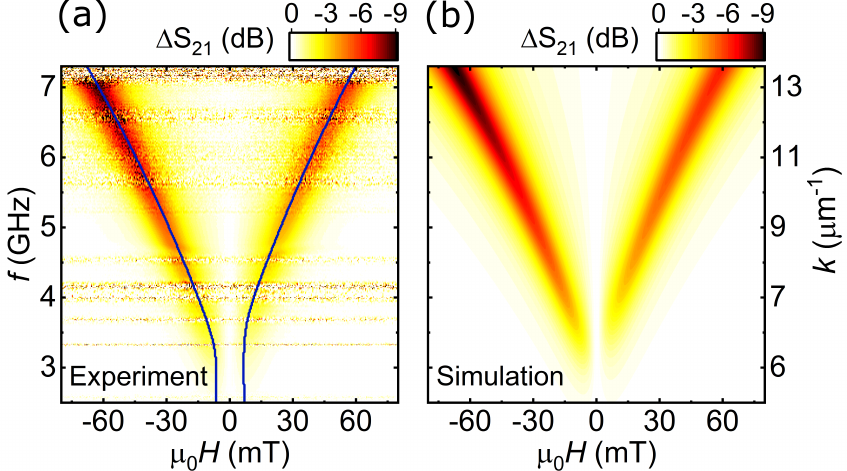}
\caption{
\label{fig:4}
(a) Nonreciprocal MASW detected in a continuous frequency range of 3 to 7~GHz. The results are shown for CoFeB(2)/Pt and $\phi_\text{H} = 45^\circ$. The blue line indicates the position of the resonance fields from the simulation in (b), carried out with the parameters in TABLE SIII~\cite{Suppl:Ref}.
}
\end{figure}

According to Eq.~(\ref{eq:2}), the DMI-induced shift in resonance field should increase with increasing wavevector. To test this, we perform our SW resonance measurements over a frequency range of $3~\text{GHz}\leqq f \leqq 7~\text{GHz}$. Although the highest signal-to-noise ratio in MASW resonance spectroscopy is obtained if the frequency of the measurement corresponds exactly to the IDT resonance frequency, it is nevertheless possible to characterize the SAW transmission signal as a function of frequency due to the relatively high bandwidth of our IDTs with only three finger pairs. These experiments resemble SAW-driven broadband SW resonance experiments, where MASWs with a quasi-continuous range of wavevectors $|k|=\frac{2 \pi f}{c_\text{SAW}}$ can be excited.
Results of this type of measurement are shown in FIG.~\ref{fig:4}(a) for the CoFeB(2)/Pt sample at $\phi_\text{H} = 45^\circ$. Here the relative change of the SAW transmission $\Delta S_{21} (f, \mu_0 H)$ is obtained by subtraction of the background offset  $S_{21} (f, \mu_0 H = -200$~mT). The simulation in FIG.~\ref{fig:4}(b) is carried out with the parameters obtained from the fit in FIG.~\ref{fig:2}(b), given in supplemental TABLE SIII~\cite{Suppl:Ref}. The resonance fields of the simulation are additionally depicted by solid lines in FIG.~\ref{fig:4}(a). Again, we observe excellent agreement between experiment and theoretical model. This confirms on the one hand the linear dependence on $k$ expected from Eq.~(\ref{eq:2}) and on the other hand that the values in TABLE SIII~\cite{Suppl:Ref} can be used to describe our experiments over a frequency range of at least 3 to 7~GHz.

In conclusion, our experimental finding of  a very large nonreciprocity effect of up to 27.9~dB/mm demonstrated here validates the potential of DMI magnetoacoustic waves for the realization of acoustic diodes or SAW valves. Promising routes towards more efficient nonreciprocal SAW devices are optimizing the magnetoelasticity and the DMI strength or lowering the SW damping constant, e.g., by employing low-damping magnetoelastic Co$_{25}$Fe$_{75}$~\cite{DanielSchwienbacher.2019,LuisFlacke.2019}. 
The excellent accordance of theory and experiment demonstrates that MASW spectroscopy can be used to characterize thin magnetic films with regards to magnetic film anisotropies, magnetoelastic coupling constants, SW damping and the average DMI strength also as a function of frequency and SW wavevector. If one assumes the lithography step being the limiting factor, it should be possible to fabricate 20~nm gratings~\cite{Vieu.2000} and to probe SWs with wavevectors above 80~\textmu m$^{-1}$, which is higher than the accessible range of BLS setups that typically extend to about 25~\textmu m$^{-1}$~\cite{Sebastian.2015}. This will allow more accurate determination of DMI and opens an avenue for on-chip chiral phononics.

\begin{acknowledgments}
We thank Emeline D. S. Nysten and Hubert J. Krenner for their discussions. This work is financially supported by the German Research Foundation (DFG) via Projects No. WI 1091/21-1, AL 618/36-1, and WE5386/5-1.
\end{acknowledgments}

%


\begin{thebibliography}{54}
\bibitem [{b1v()}]{b1value_ref}%
  \BibitemOpen
  \href@noop {} {}\bibinfo {note} {The literature value for $b_1$ was extracted
  from Ref. \cite{P.G.Gowtham.2016} by dividing the magnetoelastic coupling
  -4.2~MJ/m$^3$ of the as-deposited CoFeB(2) sample by the saturation
  magnetization 1120~kA/m.}\BibitemShut {Stop}%
\end{thebibliography}
\end{document}